\documentclass[conference]{IEEEtran}

\usepackage{amsthm,amssymb,graphicx,multirow,amsmath,color,amsfonts}
\usepackage[update,prepend]{epstopdf}
\usepackage[noadjust]{cite}
\usepackage[latin1]{inputenc}
\usepackage{bbm} 
\usepackage{pdfpages}
\usepackage{tabulary}
\usepackage{multirow}
\usepackage{comment}
\usepackage{array}
\usepackage{tabularx}

\usepackage{amsmath,amssymb}

\usepackage{fancyhdr}
\usepackage{ upgreek }
\usepackage{ tipa }
\usepackage{ gensymb }
\usepackage{ tipa }
\usepackage{cite}
\usepackage{psfrag}
\usepackage{url}
\usepackage{stfloats}
\usepackage{amsmath}
\usepackage{array}
\usepackage{fancyhdr}
\usepackage{epsfig}
\usepackage{amssymb}
\usepackage{color}
\usepackage{subfig}
\usepackage{setspace}
\usepackage{caption}
\usepackage{booktabs}
\usepackage{ dsfont }
\usepackage{algpseudocode}
\usepackage[square,numbers,sort&compress]{natbib}
\usepackage{algorithm}  
\usepackage{algorithmicx}
\usepackage{multirow}
\usepackage{comment}
\DeclareMathOperator*{\argmax}{argmax} 
\ifCLASSINFOpdf
\graphicspath{{"./pictures/"}}
\else
 \usepackage{enumerate}
 \usepackage{amsfonts,amssymb}
\fi

\IEEEoverridecommandlockouts
\begin{document}
\title{ 5G Air-to-Ground Network Design and Optimization: A Deep Learning Approach \vspace{-0.01cm}}

\author{\IEEEauthorblockN{Yun Chen$^1$, Xingqin Lin$^2$, Talha Khan$^2$, Mehrnaz Afshang$^2$, and Mohammad Mozaffari$^2$}
	\IEEEauthorblockA{ \\
		$^1$The University of Texas at Austin, USA, Email: yunchen@utexas.edu \vspace{0.1cm}\\
		$^2$Ericsson Research, Santa Clara, CA, USA,  \\Emails: \{xingqin.lin, talha.khan, mehrnaz.afshang, mohammad.mozaffari\}@ericsson.com
	}\thanks{This work was done while Yun Chen was with Ericsson Research, USA.} 
\vspace{-0.5cm}}
\maketitle

\begin{abstract}

Direct air-to-ground (A2G) communications leveraging the fifth-generation (5G) new radio (NR) can provide high-speed broadband in-flight connectivity to aircraft in the sky. 
A2G network deployment entails optimizing various design parameters such as inter-site distances, number of sectors per site, and the up-tilt angles of sector antennas.
The system-level design guidelines in the existing work on A2G network are rather limited. 
In this paper, 
 a novel deep learning-based framework is proposed for efficient design and optimization of a 5G A2G network. The devised architecture comprises two deep neural networks (DNNs): the first DNN is used for approximating the 5G A2G network behavior in terms of user throughput, and the second DNN is developed as a function optimizer to find the throughput-optimal deployment parameters including antenna up-tilt angles and inter-site distances. 
Simulation results are provided to validate the proposed model and reveal system-level design insights.
\end{abstract}

%
\IEEEpeerreviewmaketitle

\section{Introduction}
Despite phenomenal advances in terrestrial mobile communications, providing broadband in-flight connectivity (IFC) to aircraft passengers remains a pain point \cite{rula2018mile}. IFC services 
can be provided using satellites or direct air-to-ground (A2G) communications. 
Satellites have the advantage of a global coverage spanning both land and sea, which makes them suitable for intercontinental flights.
Satellite-based IFC services, however, suffer from  a limited system capacity and long latency  \cite{krichene2015aeronautical}.
The alternative approach based on A2G communications leverages cellular technology to establish direct connectivity between terrestrial base-stations (BSs) and aircraft \cite{lin2020sky, afonso2016cellular, TutorialMO}. 
 For instance, the European Aviation Network connects the European skies using satellites in combination with an A2G network based on long-term evolution (LTE) \cite{lin2020sky}.
 The Gogo Biz A2G network uses a variant of the third-generation (3G) code division multiple access (CDMA) 2000 technology to provide IFC in North America \cite{lin2020sky}. 
 A2G networks offer a larger system capacity and shorter latencies than satellites \cite{lin2020sky} but their coverage is limited to over land or along coastal belts. 
 Therefore, IFC solutions based on A2G and satellite communications complement each other.
 
 The advent of 5G new radio (NR) offers new opportunities to enhance A2G performance. 
 The existing A2G systems are based on older generations and suffer from a limited system capacity, resulting in lower data rates for the end users \cite{lin2020sky}.  
 An NR-based A2G network can benefit from
the large bandwidth, increased spectral efficiency, 
low latency, advanced antenna technologies, and ultra-lean design attributes of NR \cite{xinNR2019}.
 With this motivation, the 3rd generation partnership project (3GPP) has been developing specifications for NR-based non-terrestrial networks which include support for A2G communications.


There are several differences between an A2G network and a conventional cellular network although both employ terrestrial BSs.  
First, the traditional BSs use down-tilted antennas to serve terrestrial users whereas A2G BSs use up-tilted antennas to face the sky. Second, A2G networks typically have larger inter-site distances (ISD) than the traditional networks due to sparsely dispersed traffic demand. Third, the high aircraft speed in A2G networks poses stringent requirements on mobility. 
Therefore, the system-level design insights 
inherited from 
traditional terrestrial cellular networks are not directly applicable to A2G networks.

A2G network deployment entails optimizing various design parameters including the ISD, the number of sectors per BS site, and  up-tilt angles for sector antennas. The optimal design  depends on various factors such as traffic profile, altitude range, interference characteristics and cell load. The system design guidelines are rather obscure in the existing literature on A2G networks 
\cite{tadayon2016inflight,vondra2017performance, liu2017performance,lin2020sky}. In \cite{tadayon2016inflight}, several possibilities were outlined to enhance existing LTE systems for A2G communications. In \cite{vondra2017performance},
a performance comparison for IFC was presented for a system based on A2G and satellite communications.  
In \cite{liu2017performance}, preliminary simulations results were provided for an NR-based A2G system. In \cite{lin2020sky}, the technical issues facing an NR A2G system were discussed and its performance was evaluated in a range of frequency bands. In short, the prior work has largely been limited to A2G performance evaluation under different scenarios. This motivates the need for 
A2G network deployment optimization  to enable broadband connectivity in the sky.

The problem of A2G network optimization with various interdependent parameters is challenging and requires efficient solutions. 
A promising approach for solving complex optimization problems is to exploit tools from deep learning \cite{caldwell2018deep,sun2017learning,yu2019deep}.  
In \cite{caldwell2018deep}, an optimization framework that combined evolutionary search with deep neural networks (DNNs) was proposed for solving 
optimization problems within the class of the maximum satisfiability problem.
A power allocation problem to maximize the signal-to-interference-plus-noise-ratio (SINR) was considered in \cite{sun2017learning}, where the input and output of a resource allocation algorithm  was treated as an unknown non-linear mapping and a fully-connected DNN was used to approximate it. In \cite{yu2019deep}, 
 a set of network features was identified via deep learning for a link scheduling problem. The final solution was obtained using branch and bound or dynamic programming methods. These methods are only suitable for specific models that rely on traditional optimization algorithms. 

The main contribution of this paper is a novel deep learning-based framework developed for A2G network design and deployment optimization. 
 We propose a \emph{bi-DNN} architecture consisting of two DNNs to model the A2G network behavior and solve network optimization problems involving numerous parameters. Specifically, the first DNN is trained to approximate the user throughput in an A2G network by emulating a system-level A2G network simulator.
 The second DNN is designed to optimize the network design parameters including the antenna up-tilt angles and ISD. 
 We provide system-level simulation results to validate our model and evaluate the downlink (DL) performance of an NR A2G network in terms of user throughput and SINR. 
The results reveal useful insights on  
system performance under different antenna configurations, ISDs, number of sectors, and traffic loads. 

\section{System Model}
\begin{figure}[!t]
	\centering
	\includegraphics[width=8.8cm]{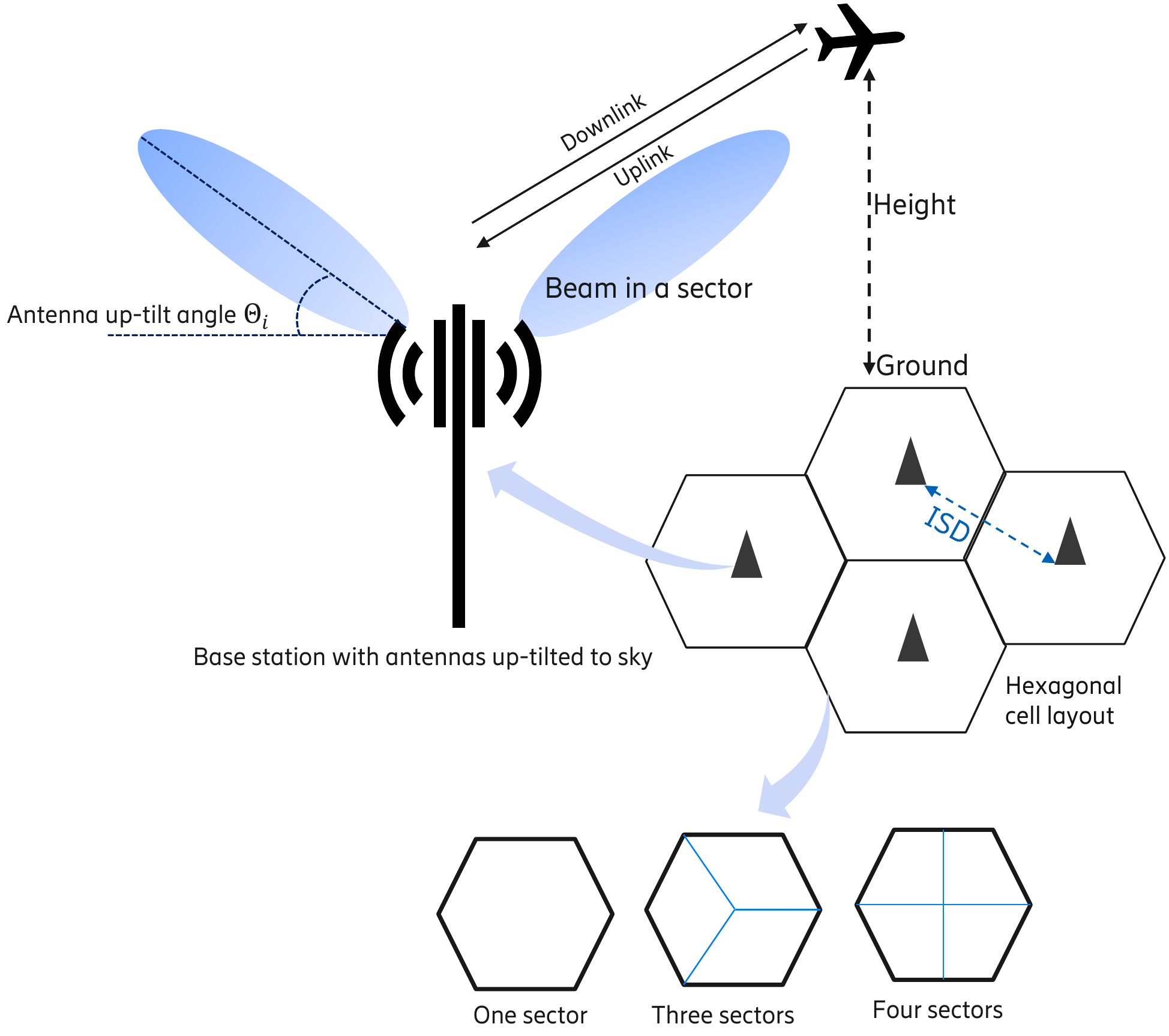}\vspace{0.2cm}
	\caption{\small Network model for A2G communications.}
	\label{SystemModel_main}
\end{figure}

We consider a dedicated A2G network consisting of a hexagonal tessellation of terrestrial BSs that exclusively serve multiple airborne aircraft within a two-dimensional plane at a certain altitude from the ground. We assume that all BSs are identical and each BS has $\textrm{S}$ sectors where the antenna for sector $i\in\{0,\cdots,\textrm{S}-1\}$ is up-tilted with an angle $\Theta_i\in[\Theta_{\min}, \,\,\, \Theta_{\max}]$ with  $\Theta_{\min}, \Theta_{\max}\in[0\degree, \,\,\, 90\degree]$. We define $\mathbf{\Theta}=[\Theta_0,\cdots,\Theta_{\textrm{S}-1}]$. We note that $\Theta_i=90\degree$ means that sector $i$'s antenna boresight is pointed upward. All sectors are otherwise identical in terms of the antenna beam pattern and transmit power. The horizontal (azimuthal) plane is split uniformly among the sectors (see Fig. \ref{SystemModel_main}).
 At a given BS, the up-tilt angle for each sector can be set independently. Given the symmetrical tessellation, we assume that the same up-tilt angle configuration $\mathbf{\Theta}$ applies to all BSs. 
We focus on three practically relevant cases for the number of sectors per site, namely $\textrm{S}\in\{1, 3, 4\}$. 
 We use $d$ to denote the ISD of the hexagonal tessellation.
We consider the 3GPP rural macrocell spatial channel model specialized to line-of-sight propagation conditions. We assume full frequency reuse such that the received DL signal at an aircraft is subjected to thermal noise as well as inter-cell interference. We summarize the notation used in this paper in Table \ref{def_notate}. 

\begin{table}[!t]
	\caption{\small Summary of notations.\vspace{-0.1cm}}
\resizebox{\columnwidth}{!}{
\begin{tabular}{@{}lp{8cm}@{}}
\toprule
Notation          & Definition                                                                            \\ \midrule
$\textrm{S}$         & Number of sectors per site                                                          \\
$t_X$               & $X$th-percentile user throughput                                                   \\
$l$               & DL higher layer traffic load                                                                 \\
$d$               & ISD                                                             \\
{$\mathbf{\Theta}$}          & $1\times\textrm{S}$ vector of antenna up-tilt angles                                                         \\
$\mathcal{N}_A$   & Approximator-NN approximates user throughput with $d$, $\mathbf{\Theta}$ and $l$ as inputs          \\
$\mathcal{N}_D$   & Optimizer-NN outputs a batch of throughput-optimal ISD and antenna configurations          \\
$\textbf{D}$      & ${n\times (\textrm{S}+2-c)}$ output from 
$\mathcal{N}_D$                                        \\
$n$               & Number of deployments in a batch output from $\mathcal{N}_D$                                 \\
$c$               & Number of fixed parameters                                                               \\
$\tau$            & Number of training steps per epoch for training $\mathcal{N}_D$                          \\
$\epsilon_{\min}$ & Minimum error for each epoch when training $\mathcal{N}_D$               \\
$\epsilon^*$      & Minimum error across all epochs when training $\mathcal{N}_D$       \\
$k$               & Number of iterations for which $\epsilon^*$ remains unchanged                                            \\
$k_{\max}$        & Maximum value for $k$ until the training for $\mathcal{N}_D$ stops\\ 
\bottomrule
\end{tabular}}
\label{def_notate}
\end{table}

   We aim to find the A2G network deployment configuration in terms of the ISD and up-tilt angles that maximizes the user throughput under the given set of system parameters. 
   The user throughput is a random quantity that depends on the SINR which itself is a function of various factors such as antenna configuration, channel gain, user distribution, network geometry and traffic load. 
 We consider the $X$-th percentile user throughput $t_X$ in our objective function. For example, half of the aircraft in a cell experience a throughput of at least $t_{50}$. 
 Specifically, we treat the optimization problem: 
 

 \begin{align} \label{OPT}
\begin{array}{rrclcl}
\displaystyle \max_{d,\mathbf{\Theta}, l} & t_X \\
\textrm{s.t.} & \displaystyle l_{\min} \le &l_{} &{}  \\
&\displaystyle d_{\min} \le & d&\le d_{\max} \\
&\displaystyle  \Theta_{\min}  \le& \mathbf {\Theta} &\le \Theta_{\max} \\
\end{array}
\end{align}
 where $l\geq l_{\min}$ is the DL traffic load which captures the higher-layer traffic demand in the network. The traffic load is a proxy for the average physical layer resource utilization in a cell. A higher traffic load results in a higher resource utilization, increasing the inter-cell interference. We note that the constraints $d_{\min}$ and $d_{\max}$ on ISD are informed by economic factors such as capital and operating expenses for a site. In our model, we set $\Theta_{\min}=0\degree$ and $\Theta_{\max}=90\degree$. 
 
The optimization problem in (\ref{OPT}) is non-convex and nonlinear. The objective function depends on a myriad of interdependent underlying factors which further compounds the problem. In practical scenarios, finding a tractable expression for the considered objective function is usually infeasible. Consequently, one has to resort to exhaustive system-level simulations which require time and compute resources. This calls for an efficient technique for modeling and solving the considered optimization problem for A2G networks. 
 

\section{Bi-DNN Based Network Deployment Optimization}
\begin{figure}[!t]
\centering
\includegraphics[width=0.9\columnwidth]{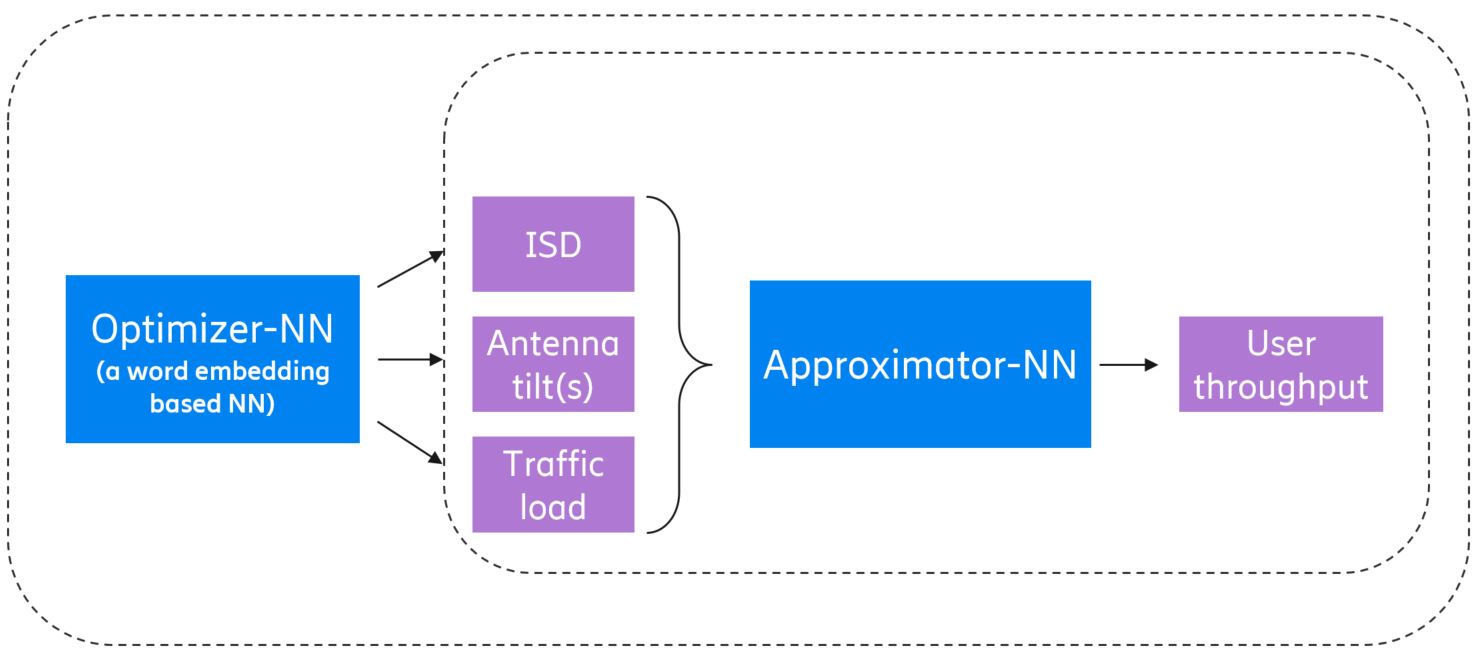}\vspace{0.1cm}
\caption{\small bi-DNN architecture for A2G network deployment optimization.}
\label{app1_model}
\end{figure}

In this section, we describe the proposed bi-DNN  architecture for A2G network deployment optimization. The bi-DNN model illustrated in Fig.~\ref{app1_model} consists of two DNNs, namely, \emph{Approximator-NN} and \emph{Optimizer-NN}. 



\subsection{Approximator-NN}
\begin{figure*}[!t]
\centering
 \subfloat[Detailed architecture of Approximator-NN, which takes the output from Optimizer-NN as input.]{%
\label{whole_model_app1}
 \includegraphics[width=\textwidth]{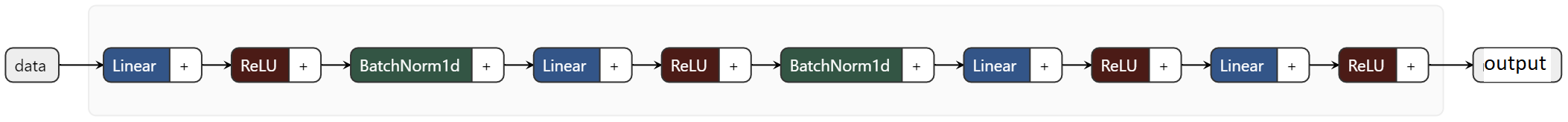}}\hfill
 \subfloat[Detailed architecture of Optimizer-NN, which outputs the settings for network parameters that serve as input for Approximator-NN.]{%
\label{model_deploy_whole}
  \includegraphics[width=\textwidth]{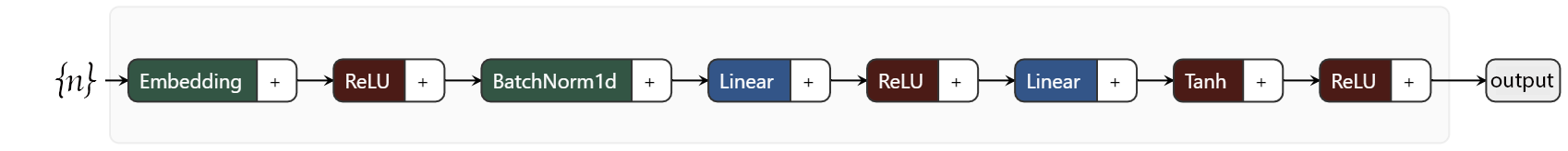}}\hfill
  \caption{\small  Detailed architectures of Approximator-NN and Optimizer-NN.}
\label{model_arc_biDNN}
\end{figure*}

Our goal is to replace the cumbersome system-level NR simulator with an agile entity to expedite the training process for Optimizer-NN.
Approximator-NN, denoted as $\mathcal{N}_{A}(\cdot)$, is a DNN designed to model the behavior of an NR A2G network and estimate its performance. 
Specifically, it approximates the $X$th-percentile user throughput for a given set of input parameters consisting of ISD, antenna up-tilt angles and traffic load. All other system parameters such as the number of BSs and the number of sectors per site remain fixed. We use the data generated from a 5G NR system-level simulator for the training and validation of Approximator-NN. Formally,
\begin{align}
    t_X &= \mathcal{N}_{A}(d,\mathbf {\Theta}, l;\textbf{w}_A)
\end{align}
denotes the approximated user throughput output by Approximator-NN, and ${\textbf{w}_A}$ denotes the neural network (NN) parameters which are initialized randomly. 
As illustrated in Fig.~\ref{whole_model_app1}, the DNN consists of 3 fully connected hidden layers with a rectified linear unit (ReLU) \cite{glorot2011deep} as the activation function. For training, we select the mean squared error (MSE) based on the throughput outputs of $\mathcal{N}_{A}(\cdot)$ and the NR simulator as the loss function,  
and use Adam \cite{kingma2014adam} as the optimizer. 

\subsection{Optimizer-NN}
Optimizer-NN, denoted as $\mathcal{N}_D(\cdot)$, is a DNN designed for solving A2G deployment optimization problem leveraging the results obtained by Approximator-NN. 
It outputs the throughput-optimal values for the A2G deployment parameters which are also fed as an input to Approximator-NN. A neural network is usually trained to minimize a certain error function. For Optimizer-NN, we set the error function equal to the negative of the average throughput of a batch of candidate deployment configurations. 
Consequently, minimizing the considered error function during the training process ends up maximizing the average throughput in the batch. 

In this work, we implement Optimizer-NN using a word embedding algorithm \cite{mikolov2013efficient}. Word embedding, usually used for solving natural language problems, aims to map the words in a word set to numerical vectors that represent the word features. In our model, we treat a batch consisting of $n$ candidate deployment configurations $\{[d_i,\mathbf{\Theta}_i,l_i]\}_{i=1}^{n}$ as the word set to be embedded, i.e., the embedding vector for the $i$th word contains the features $[d_i,\mathbf{\Theta}_i,l_i]$, same as the configuration parameters to be optimized. Hence, finding the optimal embedding vectors for the deployment configurations leads to throughput-optimal ISD and up-tilt angles.

The training process of Optimizer-NN is based on a modified version of \emph{word2vec} \cite{mikolov2013efficient}. The detailed architecture is shown in Fig.~\ref{model_deploy_whole}. A \emph{Tanh} layer is put before ReLU to normalize the output range from the network to [-1,1] so that after ReLU, the output is restricted to [0,1]. This enables proper processing of the outputs which further serve as inputs for Approximator-NN. After the convergence, the output from  Optimizer-NN consists of a batch of well-tuned configurations, among which we can select the optimal deployment that maximizes the user throughput. 
 
We next describe the modified version of word2vec algorithm for our use case. Traditionally, in a word embedding NN, the indices $\{n\} = \{0,...,n-1\}$ of the words in a given word set with cardinality $n$ constitute the input for the embedding layer. The desired outputs are the numerical embedding vectors representing either explicit or implicit features of each word in the word set. 
In this work, we  treat a set of $n$ candidate configurations as a word set, i.e., Optimizer-NN takes the set of indices $\{n\}$ as an input,  and outputs a matrix $\textbf{D}$ containing vectors representing each deployment configuration: 
\begin{align}
    \textbf{D} &= \mathcal{N}_D(\{n\}; \textbf{w}_D)
\end{align}
where the {matrix} $\textbf{w}_D$ contains the NN parameters. Furthermore, 
$\textbf{D}=\left[D_0,\cdots,D_{n-1}\right]^\textrm{T}\in\mathds{R}^{n\times (\textrm{S}+2-c)}$,
where each row $D_i$ for $i\in \{n\}$ equals  $[d_i,\mathbf {\Theta}_i,l_i]$ when $c=0$, and $\textrm{T}$ denotes the matrix transpose.
The deployment configuration $D_i$ that, when input to Approximator-NN, yields the maximum user throughput 
is chosen as the optimal solution, i.e., 
\begin{align}
\label{max_D}
D^* &= \argmax_{D_i, i\in \{n\}}\mathcal{N}_{A}({D}_i;\textbf{w}_A).
\end{align}

For training Optimizer-NN, we define the loss function $L(\mathbf{w}_D)$ as the negative of the batch average of the approximated user throughput, i.e.,
\begin{align}
    L(\mathbf{w}_D) &=\frac{-\sum_{i=0}^{n-1}t_{X,i}}{n}= \frac{-\sum_{i=0}^{n-1}\mathcal{N}_{A}(D_i;\textbf{w}_A)}{n}.
\end{align}
We use the Adam optimizer \cite{kingma2014adam} for minimizing $L(\mathbf{w}_D)$ such that the output $\textbf{D}$ from $\mathcal{N}_D(\{n\};\textbf{w}_D)$ contains candidate configurations with potentially high throughputs. We describe the training process in Algorithm \ref{DeployNN_algo}. 
We initialize the matrices $\mathbf{w}_D$ and $\textbf{D}$ randomly and run $\tau$ iterations per epoch (lines \ref{experi_begin}-\ref{experi_end}). If we intend to fix (rather than optimize) any of the parameters  $\{d,\mathbf{\Theta}, l\}$, the fixed parameter(s) can be concatenated with the output from Optimizer-NN (line 8).
We compute the mean error for the loss function in line \ref{mean_err}.
The batch $\textbf{D}'$ with the minimum error is stored for every epoch (lines \ref{step_err_begin}-\ref{step_err_end}) and the batch $\textbf{D}^*$ with the minimum error across all the epochs $\epsilon^*$ is retained (lines \ref{epo_err_begin}-\ref{epo_err_end}).  Optimizer-NN converges when $\epsilon^*$ remains unchanged for more than $k_{\max}$ iterations. Finally, the throughput-optimal parameter values are obtained using (\ref{max_D}) (line \ref{D_best}). The algorithm has a linear time complexity, 
which makes it attractive for implementation.  
\begin{algorithm}[!t]  
  \caption{ : Optimizer-NN training procedure}  
  \label{DeployNN_algo}  
  {\footnotesize
  \begin{algorithmic}[1]   
  \State Initialization:
  \Statex Choose $\textrm{S}$ and $c$
  and find embedding size: $\textrm{S} + 2-c$;
  \Statex Choose the batch size $n$ and parameters $\tau$ and $k_{\max}$; 
  \Statex  Set $\epsilon_{\min} = 0$, $\epsilon^* = 0$ and $k=0$; 
 \While {$k<k_{\max}$} \label{ear_stop}
 \State $\epsilon_{\min} = 0$;
 \For{$\tau_i=1:\tau$} \label{experi_begin}
 \State $\textbf{D}=\mathcal{N}_D(\{n\};\textbf{w}_D)\in \mathds{R}^{n\times (\textrm{S}+2-c)}$;
 \State $\textbf{D}'\gets \textbf{D}$;
 \If{$c>0$}
 \State Construct $\textbf{D}'\in\mathds{R}^{n\times (\textrm{S}+2)}$ by concatenating the $c$ fixed  
 \Statex \qquad\qquad\, parameters to $\textbf{D}$; 
 \EndIf
 \State Calculate error  $\upepsilon=-\mathcal{N}_A(\textbf{D}';\textbf{w}_A)\in \mathds{R}^{n\times 1}$;
\State Evaluate loss function: $L(\textbf{w}_D) = \text{mean}(\upepsilon)$; \label{mean_err}
 \State Optimize $L(\textbf{w}_D)$ using Adam optimizer;
 \If{$\min(\upepsilon)< \epsilon_{\min}$}\label{step_err_begin}
 \State $\textbf{D}^{\dag}\leftarrow \textbf{D}'$;
 \State $\epsilon_{\min} \leftarrow \min(\upepsilon)$;
 \EndIf \label{step_err_end}
 \EndFor\label{experi_end}
 \If{$\epsilon_{\min}< \epsilon^*$} \label{epo_err_begin}
 \State $\epsilon^* \leftarrow \epsilon_{\min}$;
 \State $\textbf{D}^*\leftarrow \textbf{D}^{\dag}$;
 \State $k\leftarrow 0$;
 \Else
 \State $k = k + 1$;
 \EndIf \label{epo_err_end}
 \EndWhile
 \State $D^* = \argmax\limits_{D_i,i\in\{n\}}\mathcal{N}_A(\textbf{D}^*;\mathbf{w}_A)$; \label{D_best}
\end{algorithmic}}
\end{algorithm}

\section{Simulation Results and Discussions}
We generate the dataset for our simulations using a system-level simulator for 5G NR. The simulator models the DL of an A2G network consisting of 19 BSs serving aircraft located randomly within the simulation area. The BS height is 35 m while the aircraft are at an altitude of 12 km. For each sector, we assume a transmit power of 49 dBm and a cross-polarized $4\times4$ planar array with half-wavelength element spacing.
We set the carrier frequency and system bandwidth to 3.5 GHz and 100 MHz. We assume full frequency reuse and consider wrap-around to account for inter-cell interference.
We discuss the simulation results for Approximator-NN and Optimizer-NN in section \ref{sec: training} and \ref{sec: deploy}. For performance comparison, we use the optimal deployment configuration obtained from an exhaustive search of the dataset as a benchmark. 

 
\subsection{Approximator-NN training and validation}\label{sec: training}
We train Approximator-NN to output the 50th percentile user throughput $t_{50}$. We use the parameter values shown in Table \ref{ds_app1} for data collection. 
For $S=1$, the number of neurons in each hidden layer is $10$. For $S\in\{3,4\}$, each hidden layer has $16$ neurons to further improve the approximation accuracy. 
The training dataset consists of  $760$, $48905$, and $30405$ entries for $S=1$, $S=3$, and $S=4$, respectively. 
\begin{table}[]
	\caption{\small Dataset parameter ranges for Approximator-NN.}
\resizebox{\columnwidth}{!}{
\begin{tabular}{|l|l|l|}
\hline
Paramter        & Range                          & Step size                               \\ \hline
ISD             & {[}20, 160{]} km               & 20 km                                   \\ \hline
Up-tilt angles & {[}0$\degree$, 90$\degree${]} &  5$\degree$ ($S\in\{1,3\}$); 10$\degree$ ($S=4$)\\ \hline
Traffic load         & {[}10, 70{]} Mbps              & 20 Mbps                                 \\ \hline
\end{tabular}}
\label{ds_app1} 
\end{table}

\begin{figure}[!t]
\centering
\includegraphics[width=.9\columnwidth]{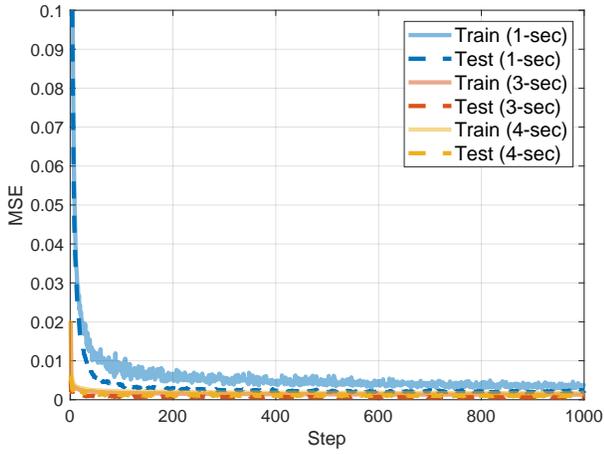}
\caption{\small Convergence of Approximator-NN.}
\label{conver_all}
\end{figure}

\begin{figure}[!t]
	\centering
		\includegraphics[width=0.85\columnwidth]{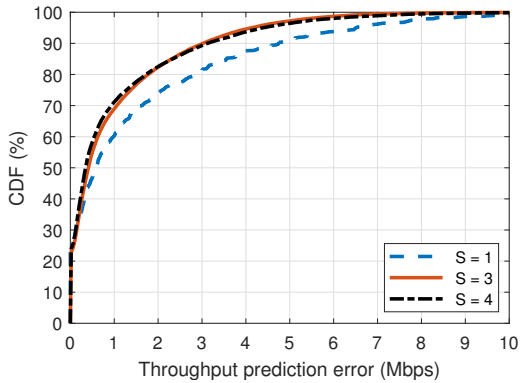}\hfill
	\caption{\small Approximator-NN approximation performance:  absolute prediction errors using the data in the dataset.}
	\label{AxPreErr}
\end{figure}

In Fig.~\ref{conver_all}, we show the convergence of the training process in terms of the MSE of the user throughput. 
In Fig.~\ref{AxPreErr}, we plot the cumulative distribution function (CDF) of the user throughput prediction error. We observe that for $S=1$, 95\% of the prediction errors are 6.5 Mbps or lower. For $S\in\{3,4\}$, 95\% of the prediction errors are within 4.5 Mbps thanks to
{larger datasets}. 
We emphasize that our goal is to optimize A2G network parameters to achieve high throughput, which is typically on the order of hundreds of Mbps in the considered scenario. This suggests that the throughput approximation errors in Approximator-NN are relatively small compared to the anticipated throughput range of an A2G system.  
Therefore, Approximator-NN can provide satisfactory throughput approximations for a given input $[d, \mathbf {\Theta}, l]$.

\subsection{A2G network deployment optimization}\label{sec: deploy}

\begin{figure}[!t]
	\centering
	\includegraphics[width=0.8\columnwidth]{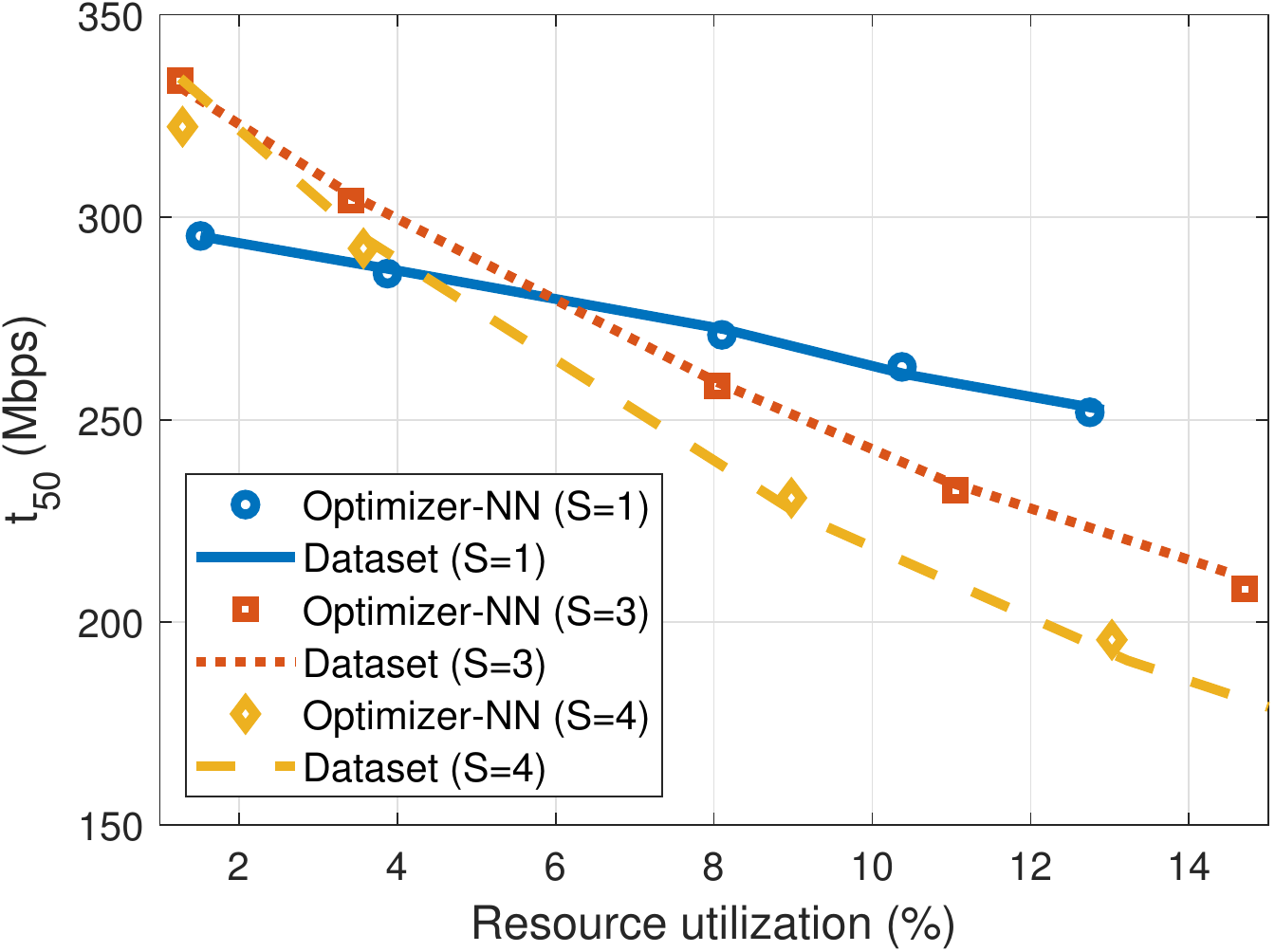}\hfill
	\caption{\small User throughput under varying load conditions for the optimal configurations based on Optimizer-NN and the dataset.}
	\label{NoFix_comb_tpt}
\end{figure}

\begin{figure}[!t]
	\centering
	\includegraphics[width=0.8\columnwidth]{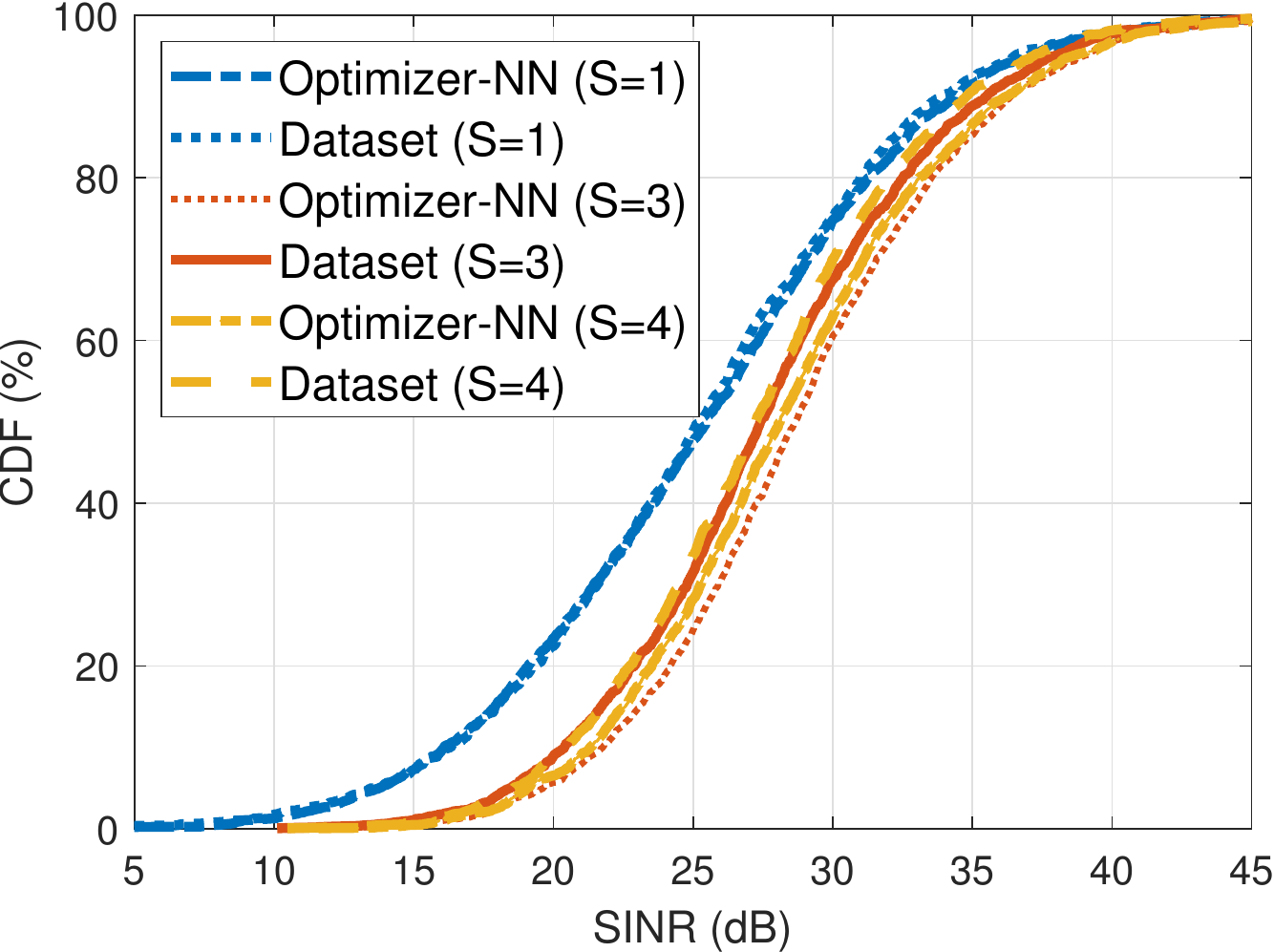}\hfill
	\caption{\small SINR CDF for the optimal configurations based on Optimizer-NN and dataset for $S\in\{1,3,4\}$.}
	\label{NoFix_comb_SINR}
\end{figure}


We obtain optimal deployment configuration using Algorithm \ref{DeployNN_algo} with parameter values $\tau=1000$, $k_{\max}=20$, $n=4$ for $S=1$, and $n=8$ for $S\in\{3,4\}$.
For each $S\in\{1,3,4\}$, the optimal configuration based on Optimizer-NN has an ISD of 20 km and a traffic load of 4 Mbps, same as the minimum allowed values  in (\ref{OPT}). The optimal configurations in the dataset have an ISD of 20 km and a traffic load of 10 Mbps, same as the minimum possible values in the dataset. This trend is plausible since a smaller ISD helps improve the received signal power while a lower traffic load reduces the inter-cell interference. 

In Table \ref{NoFix_tpt_table}, we provide the optimal values for the antenna up-tilt angles with the corresponding user throughput for both Optimizer-NN and the dataset. The reported results for throughput are obtained directly from the simulator by plugging in the optimized configuration parameters.
We find that the antennas need to be up-tilted somewhat aggressively (around 60$\degree$ to 90$\degree$) due to a relatively small ISD of 20 km. For the 4-sector case, one antenna needs to be in an almost upright position to boost throughput. This is because three sectors are sufficient to cover the entire cell due to its small size. With an almost upright fourth sector, the inter-cell interference is reduced as more power is directed to the region above the BS.

We note that the proposed algorithm learns the benefit of operating in a lightly-loaded system, as evident from the optimal load of 4 Mbps based on Optimizer-NN despite the minimum value of 10 Mbps considered in the dataset. Consequently, the throughput for the optimal configurations based on Optimizer-NN is higher than that based on the dataset.

We next consider another practically relevant scenario where the ISD is large. In Table \ref{ISD80_tpt_table}, we provide the optimal antenna up-tilt angles for different number of sectors for a fixed ISD of 80 km. We observe that due to a larger ISD, it is optimal to have only moderately up-tilted antennas (around $30\degree$) 
for coverage extension. This finding holds for all considered values for the number of sectors. 
With a small ISD, the network is interference-limited and using highly up-tilted antennas (as in Table \ref{NoFix_tpt_table}) helps direct the signal power towards the aerial region above the BS, thus reducing inter-cell interference. As the cell size expands, the network becomes coverage-limited and the antennas need to be slanted to provide coverage throughout the cell.  
Furthermore, the larger ISD also results in  a lower throughput. For instance, compared to the case with 20 km ISD, $t_{50}$ is reduced by 49\%, 35\% and 32\%  for $S=1$, $S=3$, and $S=4$, respectively.
 
In Fig.~\ref{NoFix_comb_tpt}, we examine the throughput variation for the optimal antenna configuration in Table \ref{NoFix_tpt_table} and the optimal ISD under different traffic conditions. A higher traffic load corresponds to a higher resource utilization. The plot reaffirms the intuition that a higher traffic load leads to a stronger inter-cell interference, thus reducing the average user throughput. We also observe that the throughput for the optimal configuration obtained using the proposed framework is almost indistinguishable from that obtained using the dataset. Another noticeable trend in Fig.~\ref{NoFix_comb_tpt} is the throughput variation for different sectors under varying traffic conditions. Under a low traffic load, the 1-sector case  results in the lowest throughput among all cases. As the traffic load increases, the 1-sector case yields the highest throughput while the 4-sector case the lowest. This is because deploying more sectors per site reduces the cell size which increases the received signal power for a typical user. This, however, also increases the total radiated power per site, which manifests as inter-cell interference. Under a high traffic load, the degradation due to interference overshadows the improvement in the received signal power, thus reducing the SINR and throughput. Finally, in Fig.~\ref{NoFix_comb_SINR}, we observe similar SINR CDFs for the optimal configurations based on Optimizer-NN and the dataset.




\begin{table}[!t]
	\caption{\small Up-tilt angles based on Optimizer-NN  and  dataset for the optimal ISD $d=20$ km.}
	\resizebox{\columnwidth}{!}{
		\begin{tabular}{|l|c|c|c|c|}
			\hline
			\multirow{2}{*}{$\textrm{S}$} & 
			\multicolumn{4}{l|}{\qquad\qquad Optimizer-NN \qquad\qquad\qquad\qquad \quad  Dataset}                   \\ \cline{2-5} 
			& $\mathbf{\Theta}$                & $t_{50}$ (Mbps) & $\mathbf{\Theta}$                  & $t_{50}$ (Mbps) \\ \hline
			1                          & 78\degree                       & 295.39    & 70\degree                      & 287.27    \\ \hline
			3                          & {[}57\degree, 58\degree, 57\degree{]}      & 333.68    & {[}60\degree, 60\degree, 60\degree{]}      & 305.51    \\ \hline
			4                          & {[}90\degree, 58\degree, 62\degree, 61\degree{]} & 322.36    & {[}50\degree, 80\degree, 60\degree, 50\degree{]} & 295.35    \\ \hline
	\end{tabular} }
	\label{NoFix_tpt_table}
\end{table}

\begin{table}[!t]
	\caption{\small Up-tilt angles based on Optimizer-NN and dataset for a fixed ISD $d=80$ km.}
	\label{ISD80_tpt_table}
	\resizebox{\columnwidth}{!}{
		\begin{tabular}{|l|c|c|c|c|}
			\hline
			\multirow{2}{*}{$\textrm{S}$} & \multicolumn{4}{l|}{\qquad\qquad Optimizer-NN \qquad\qquad\qquad\qquad \quad  Dataset}                   \\ \cline{2-5} 
			& $\mathbf{\Theta}$                & $t_{50}$ (Mbps) & $\mathbf{\Theta}$                  & $t_{50}$ (Mbps) \\ \hline
			1                          & 35\degree                      & 149.07    & 30\degree                      & 138.76    \\ \hline
			3                          & {[}29\degree, 32\degree, 27\degree{]}      & 216.48    & {[}30\degree, 30\degree, 30\degree{]}      & 198.18    \\ \hline
			4                          & {[}26\degree, 35\degree, 36\degree, 37\degree{]} & 216.56    & {[}30\degree, 30\degree, 30\degree, 30\degree{]} & 195.68  \\ \hline 
	\end{tabular}}
	
\end{table}

\section{Conclusion}
In this paper, we have developed a new deep learning-based framework for A2G network design and deployment optimization. Specifically, 
we have proposed a bi-DNN architecture for modeling and optimizing A2G networks that involve a wide range of parameters. In the proposed framework, the first DNN helps approximate the aircraft throughput and the second DNN determines the throughput-optimal network design parameters including the antenna up-tilt angles and the ISD. 
We have drawn several insights from the simulation results.
It is throughput-optimal to aggressively up-tilt the antennas when the ISD is small. With a large ISD, the network gets coverage-limited and the antennas need to be slanted to provide coverage throughout the cell.
Moreover, under a low traffic load, it is throughput-optimal to deploy a large number of sectors. Under a high traffic load, the network becomes interference-limited and operating with fewer sectors is beneficial as it reduces interference.   

\footnotesize
\bibliographystyle{IEEEtran}
\bibliography{IEEEabrv, references}

\end{document}